\documentclass[prb,twocolumn,twoside,showpacs,floatfix]{revtex4}
\usepackage{graphics,amssymb,amsmath,epsf}
\epsfclipon

\begin{document}

\title{Coordination motifs and large-scale structural organization
in atomic clusters}

\author{Zhu Yang and Lei-Han Tang}
\affiliation{Department of Physics, Hong Kong Baptist University, 
Kowloon Tong, Hong Kong SAR, China}

\date{\today}

\begin{abstract}
The structure of nanoclusters is complex to describe due
to their noncrystallinity, even though bonding and packing constraints
limit the local atomic arrangements to only a few types.
A computational scheme is presented to extract coordination motifs
from sample atomic configurations. The method is based on a clustering
analysis of multipole moments for atoms in the first coodination shell.
Its power to capture large-scale structural properties 
is demonstrated by scanning through the ground state of the Lennard-Jones
and C$_{60}$ clusters collected at the Cambridge Cluster Database.

\end{abstract}

\maketitle

\section{Introduction}

Nanoscale atomic clusters have attracted a great deal of attention
in recent years due to their promising applications in catalysis, photonics, 
and bioimaging.\cite{ref:cluster-tech,ref:gold-cluster,ref:cluster-RMP}
The functionality of these 
clusters is controlled by their structure, which may assume a variety
of forms as a result of the generally complex energy 
landscape governing the low energy cluster conformations.\cite{ref:DJW-book} 
Atomic arrangements in these clusters generally do not follow the
crystal structure of the corresponding bulk material, but may
nevertheless possess certain regularities.
Suitable representation of these structures, particularly for
small and medium sized clusters, is important for uncovering
mechanisms that underly their thermodynamic stability and functionality.

Traditionally, atomic clusters of high point symmetry are described in 
terms of the five Platonic solids and their truncated 
polyhedra.\cite{ref:DJW-book,ref:Martin}
Starting from, say, a 13-atom core as illustrated in Fig. 1, 
successive layers of atoms can be added to produce clusters of ``magic size''
which feature prominently in spectroscopic measurements.
Two particularly well-known series are the Mackay icosahedra\cite{ref:Mackay}
(MIC) that grow from Fig. 1(c) and the Marks decahedra\cite{ref:Marks}
(MD) that grow from Fig. 1(d).
Noble-gas nanoclusters follow the Mackay icosahedra series up to 
a few thousand atoms before switching to the Marks decahedra 
series.\cite{ref:noble-gas-exp}
Both types of clusters can be viewed as multiply twined particles
that take advantage of the low energy surfaces of the constituting 
face-centered-cubic (fcc) grains. At even larger sizes, accumulation of 
strain in the particle drives a transition to the single crystal structure,
although the size at which such transition takes place is still 
debated.\cite{ref:LJC}

\begin{figure}
\narrowtext
\centerline{
\epsfxsize=\linewidth
\epsfbox{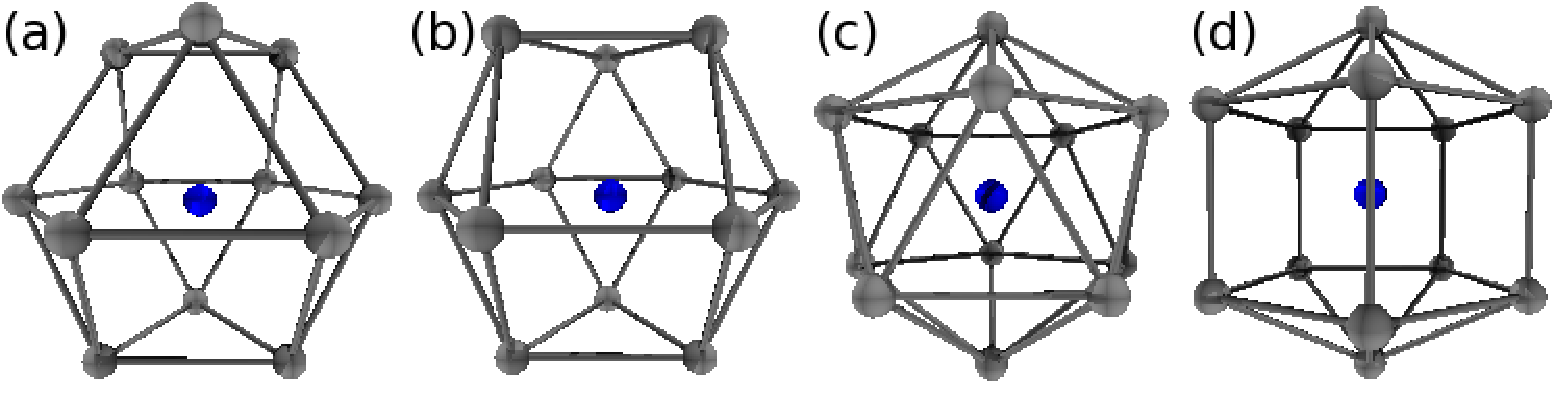}
}
\caption{(Color online) The first coordination shell at coordination number 12.
(a) face-centered-cubic (fcc); (b) hexagonal-close-packed (hcp); 
(c) icosahedral (ico); (d) decahedral (dec).}
\label{fig1}
\end{figure}

Real structures often contain defects and imperfections as compared to 
the ideal ones.\cite{ref:vandeWaal} Recently, Polak {\it et al.} 
investigated medium-sized Lennard-Jones (LJ) clusters of a few hundred
atoms obtained in Monte Carlo simulations at low 
temperatures.\cite{ref:Polak03,ref:Polak08} 
They proposed a scheme to describe
the interior structure of an atomic cluster in terms of the four
``coordination polyhedra'' of Fig. 1. Indeed, when atoms with a particular
type of coordination are displayed, large-scale features such as cubic
domains or polyicosahedral grains become readily visible.
Polak applied the scheme to describe the structure evolution 
during cooling runs, and in cluster growth starting from a seed crystallite.
It is conceivable that insights generated from such studies can assist 
experimental designs aimed at producing a particular class of nanoclusters.

The four coordination polyhedra in Fig. 1 were identified by Polak through 
visual inspection of a large number of atomic configurations.
Here we extend Polak's work to construct a systematic and
robust computational methodology for local coordination analysis and
large-scale structure identification.
This is necessary as the local atomic structures often contain distortions
or even belong to unknown categories. The idea is quite simple.
Given a set of atomic coordinates, we first compute the radial 
distribution function to determine a suitable cutoff 
distance for the first coordination shell. We then parametrize
the coordination shell configuration of each interior atom using 
the multipole moments $Q_l$ of the shell atoms.\cite{ref:SNR}
The data set is examined for their clustering properties in the 
``shape space'' spanned by the $Q_l$'s.
Each well-defined cluster in the shape space represents a re-occuring 
local structural unit which we call a ``coordination motif''.
Atoms in the same cluster are then assigned the same ``type''.
The scheme allows for fast classification of interior atoms and their
local neighborhoods. The geometric shape of individual motifs
and their spatial organization provide valuable information on 
large-scale structure organization. The MIC and the MD
series, for example, each possesses a unique composition profile of the 
coordination motifs. In such cases, the motif content alone can be used 
for accurate structure association.

The paper is organized as follows. In Sec. II we describe the computational
procedure for the identification of structure motifs. The ground state LJ 
clusters are used as examples. Section III contains a discussion of 
large-scale structure organization based on the spatial distribution of 
motifs of different types. Particular emphasis is given to the decagonal 
motifs that define local pentagonal symmetry axis for twinning.
Section VI contains a summary of our findings and conclusions from the work.

\section{Local structure motif identification}

The procedure for the coordination motif identification consists of three 
steps: i) specification of the first coordination shell through an analysis 
of the radial distribution function; ii) parametrization of the 
atomic configuration for completed shells;
iii) clustering of the data to determine the most frequently
occuring coordination motifs.

We illustrate the method in the description of 
the (putative) ground state of LJ clusters with up to $N=1610$ atoms,
available at the Cambridge Cluster Database (CCD).\cite{ref:CCD,ref:Shao}
Figure 2(a) shows the radial distribution function
obtained from the histogram of pairwise distances 
$r_{ij}=|{\bf r}_i-{\bf r}_j|$ for three selected atomic clusters 
at $N=55, 549$ (complete Mackay icosahedra) and $1200$, at a bin size
$\Delta r=0.01$ (in units of the LJ radius $\sigma$).
Note that the minimum of the LJ potential is at 
$r_{\rm min}=2^{1/6}\sigma\simeq 1.12\sigma$.
In all three cases, there are well-defined 
distance gaps which allow for unambiguous identification of first
and second coordination shells for all atoms on the cluster.
The position of the first two peaks are at $r_1\simeq 1.1$ and
$r_2\simeq 1.55$, respectively, with the ratio $r_2/r_1\simeq 1.4$ which
corresponds well to the value $\sqrt{2}$ for the fcc or hcp structure.
The plot suggests a cut-off distance $r_{\rm cut}=1.3$ for atoms to
be included in the first coordination shell. To distinguish interior from
surface atoms, we compute the offset distance $r_{\rm offset}$
between the center atom and the center of mass (CM) of its first 
coordination shell. Figure 2(b) shows a scatter plot of $r_{\rm offset}$
against the coordination number for all atoms on the $N=549$ cluster.
The approximately linear behavior of the data indicates that completed
shells have 12 atoms. 

\begin{figure}
\narrowtext
\centerline{
\epsfxsize=\linewidth
\epsfbox{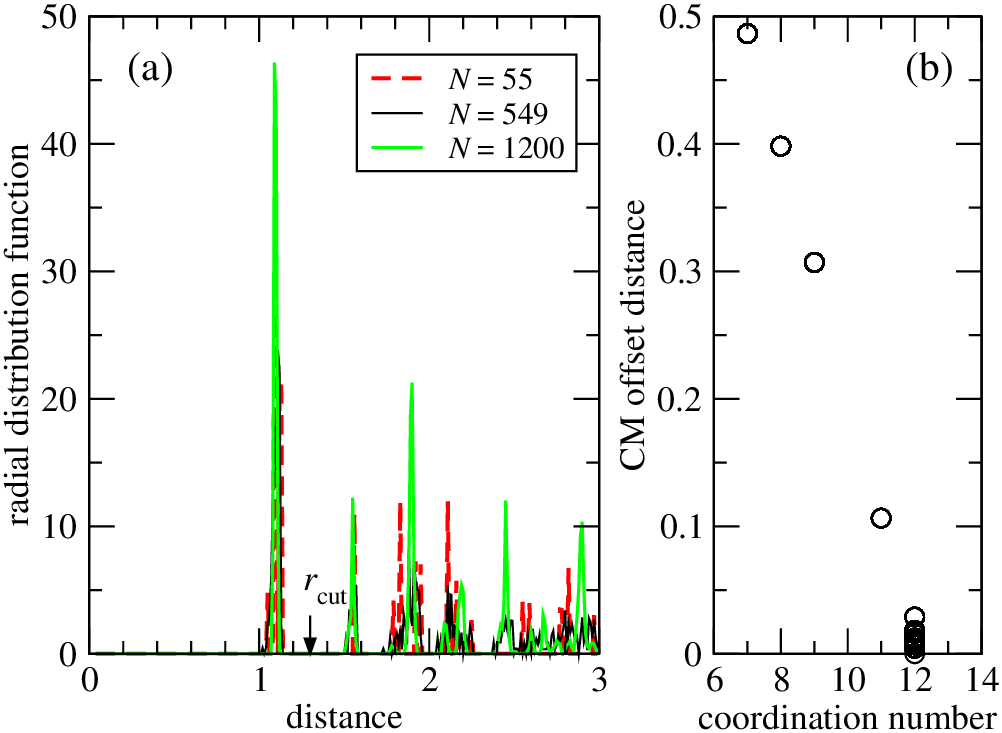}
}
\caption{(Color online) (a) Radial distribution function 
of LJ clusters at three selected sizes. 
(b) The distance between the center of mass (CM) of the first coordination
shell and the center atom against the coodination number for all atoms
in a $N=549$ cluster. Here $r_{\rm cut}=1.3$.}
\label{fig2}
\end{figure}

The 12-atom neighborhoods as shown in Fig. 1 are distinguished from
one another by the angular distribution of atoms on the shell.
Provided the distance of shell atoms to the center atom falls in a narrow 
range, the shell configuration is well represented by a density function
$\rho(\theta,\phi)=\sum_k \delta(\theta-\theta_k,\phi-\phi_k)$
on the unit sphere, where $(\theta_k,\phi_k)$ are the polar coordinates 
of shell atom $k$ with the origin on the center atom.
Consider now the ``multipole'' expansion in the spherical harmonics,
\begin{equation}
\rho(\theta,\phi)=\sum_{l,m}a_{lm}Y_{lm}(\theta,\phi),
\label{sp_expansion}
\end{equation}
where $a_{lm}=\sum_k Y_{lm}^*(\theta_k,\phi_k)$.
The multipole moments of the shell configuration are given 
by,\cite{ref:SNR}
\begin{equation}
Q_l\equiv{1\over C}\sqrt{{4\pi\over 2l+1}\sum_{m=-l}^l|a_{lm}|^2}
=\sqrt{{4\pi\over 2l+1}\sum_{m=-l}^l|\overline{Y}_{lm}^*|^2},
\label{BOO}
\end{equation}
where $C$ is the number of shell atoms, and the overline bar
denotes average over the shell.
Note that $Q_l=1$ (all $l$) if there is only one atom on the shell.
It is easy to verify that the multipole moments $Q_l$ are invariant
with respect to a rigid rotation of the shell density.

In Ref.\cite{ref:Polak03}, the moments $Q_4$ and $Q_6$ are used to 
distinguish the four atomic configurations shown in Fig. 1. We propose 
here a more general approach based on the clustering property of 
interior atoms in the shape space spanned by the multipole moments $\{Q_l\}$. 
Atoms that are close to each other in this representation have
very similar neighborhoods. Strong clustering of the data points 
in the shape space suggests existence of only a few coordination motifs.
Standard clustering methods\cite{ref:clustering} can then be used to 
assign coordination motifs to the atoms. 
On the other hand, weak clustering would imply a more 
continuous spectrum of local neighborhoods without a clear local
structural preference.

Figure 3 shows the multipole moments $Q_4$, $Q_6$ and $Q_8$ 
for all interior atoms on the ground-state LJ clusters at the three 
selected sizes. In all cases, the data exhibit very strong clustering
which agrees well to the four shell configurations in Fig. 1. 
Note that the cluster that corresponds to the decagonal motif
has a slightly bigger spread than the other three clusters.
Inspection of the atomic coordinates show that the decahedron has a 
shorter axis in the two smaller systems and is elongated slightly
in the $N=1200$ case. The moment $Q_8$ performs better than
$Q_4$ in discriminating the hcp and dec motifs. 

\begin{figure}
\narrowtext
\centerline{
\epsfxsize=\linewidth
\epsfbox{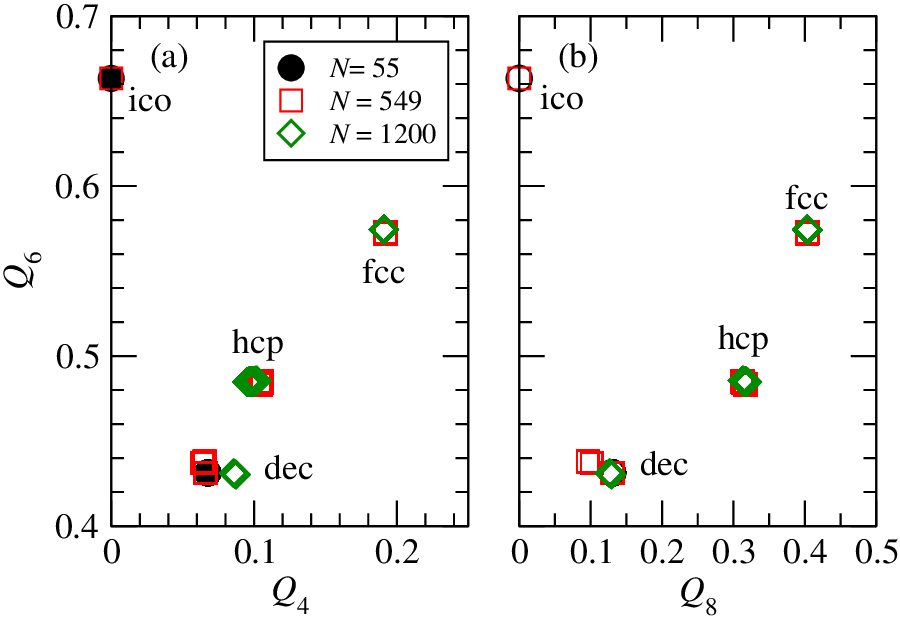}
}
\caption{(Color online) Scatter plot of the multipole moments of 
interior atoms on the ground state LJ clusters 
at three different cluster sizes.
(a) $Q_6$ versus $Q_4$; (b) $Q_6$ versus $Q_8$. 
}
\label{fig3}
\end{figure}

Thermal fluctuations lead to distortions of the polyhedra so that
the corresponding data points in the shape space are more spread
out than in the ground state.
We have investigated the dependence of the width of such clouds
on the noise strength using simulated data.
Starting from each of the four regular polyhedra, we
randomly displace the 12 shell atoms laterally
to create noisy density patterns. The displacement of each atom is
restricted to within a distance $\Delta r=R\Delta\theta$ from its 
original position, were $R$ is the radius of the coordination shell
and $\Delta\theta$ defines the noise strength.
Figures 4(a)-(c) show a comparison of the $Q_l$'s for the ideal (solid
symbols) and noisy (open symbols) structures at $\Delta\theta=0.07\pi$. 
In the latter case, 100 shell configurations are generated from each
of the four regular polyhedra. Error bars on the data 
in Fig. 4(b) represent standard deviations over the 100 samples.
From the plots it is seen that the icosahedral coordination motif
is well separated from the other three and can be easily recognized.
The hcp motif is sandwiched between the fcc and dec motifs
in the shape space and is more easily deformable into either of the two.
At $\Delta\theta=0.07\pi$, the hcp cluster 
already begin to merge with the fcc and dec clusters.
The data (with error bars) in Fig. 4(b) suggest that either $Q_6$ or $Q_7$
can be used to separate the hcp from the fcc motif while $Q_8$ can be used
to separate the hcp from the dec motif. Enhanced performance is expected
when the classification is done in the high dimensional shape space
which combines information from all the $Q_l$'s.

\begin{figure}
\narrowtext
\centerline{
\epsfxsize=\linewidth
\epsfbox{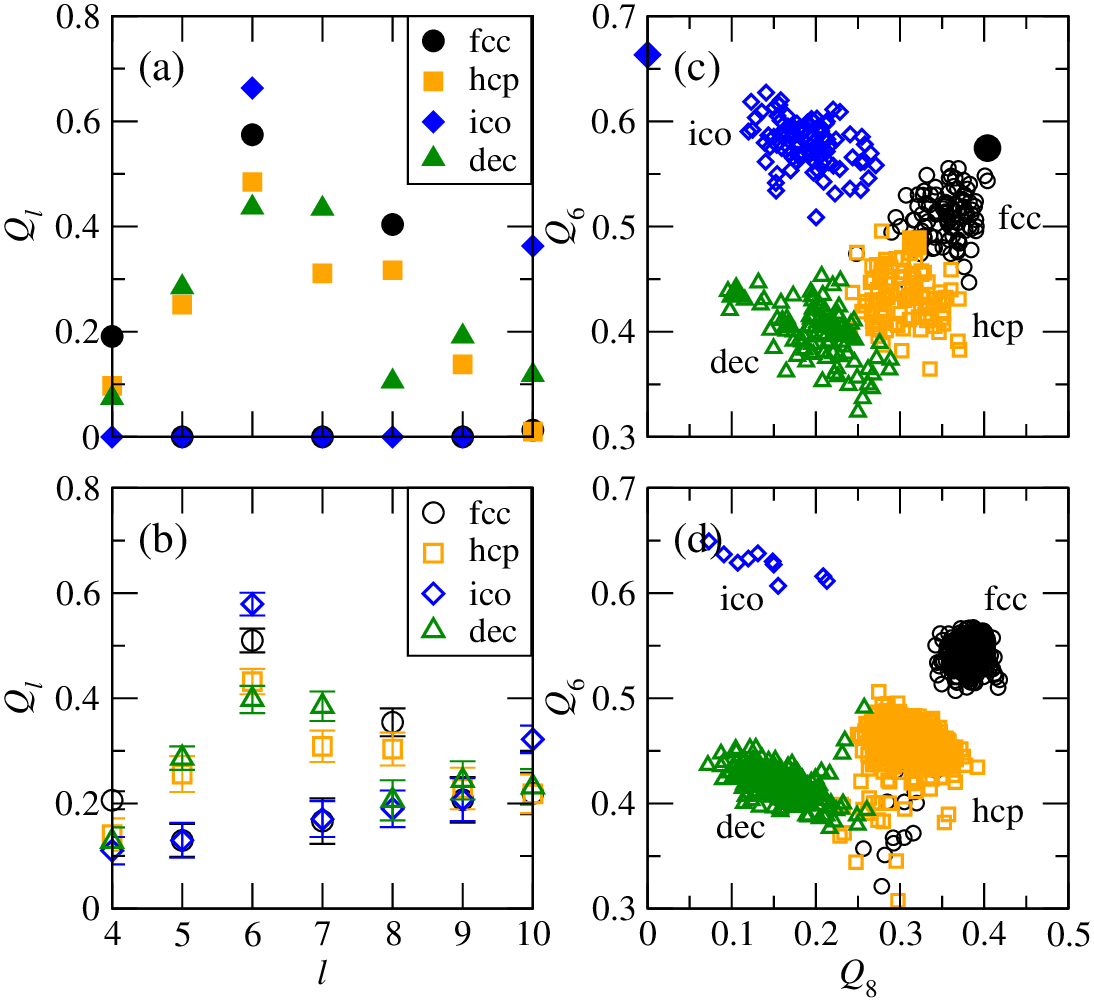}
}
\caption{(Color online) The multipole moments $Q_l$ for 
regular (solid symbols) and noisy (open symbols) polyhedra. 
(a) $Q_l$ against $l$ in the noiseless case;
(b) $Q_l$ against $l$ when shell atoms are displaced laterally at
a noise strength $\Delta\theta=0.07\pi$;
(c) $Q_6$ versus $Q_8$ for individual shell configurations
at $\Delta\theta=0.07\pi$;
(d) $Q_6$ versus $Q_8$ from 10 randomized samples of the $N=549$ 
LJ cluster at $\Delta r = 0.1 r_{\rm min}$ (see text).
}
\label{fig4}
\end{figure}

We have also generated noisy atomic clusters by adding random displacements
to the ground state LJ configurations. Each component of the atomic 
displacement is
taken to be gaussian distributed with zero mean and standard deviation $h$.
The resulting atomic cluster is analyzed following the procedure described
above. Figure 4(d) shows an example of the local coordination
data obtained from 10 randomized samples of the
$N=549$ cluster at $h=0.06 r_{\rm min}$ or 
$\Delta r=3^{1/2}h=0.1 r_{\rm min}$. Only atoms with coordination
number 12 are included. The symbols used for the selected
atoms in the plot correspond to their coordination type in the ground
state cluster. It is seen that the data at this noise level
still form well-defined clusters in the shape space, with 
the majority of atoms keep their original identity.
There are however a few outliers (points with $Q_6$ less than 0.35)
where the shell configuration has changed so significantly that
they can no longer be classified into any of the four categories.
Closer inspection of the atomic configurations
indicate that such cases are usually accompanied by large displacement
of atoms in the neighborhood so that a more careful definition of
the first coordination shell for the atom in question is needed.
Interestingly, the Lindemann criterion for crystal melting sets
a limit on $\Delta r$ to be around 10-12\% of the nearest neighbor
distance at the bulk melting point.\cite{ref:cahn} 
From our simulation studies, this is also
the noise level around which different coordination motifs start
to become indistinguishable from each other.

\section{Composition profile and large-scale structure organization
of atomic clusters}

Using the moment pair $(Q_6, Q_8)$ as classifier,
we computed the composition profile $c_\alpha=N_\alpha/N_{\rm interior}$
of the four coordination motifs 
for all ground state LJ clusters in the size range $N=13$ -- 1610. 
Here $N_\alpha$ is the number of $\alpha={\rm fcc,\ hcp,\ ico,\ dec}$ atoms
in the cluster interior and $N_{\rm interior}$ is the total number
of interior atoms. Results are shown in Fig. 5.
Each curve gives the percentage of a particular motif
in the atomic cluster against the cluster size.
As a reference, Table I lists the number of interior atoms of each
type for complete Mackay icosahedra and Marks decahedra as a function of 
the layer number $n$. Also given are the total number of interior atoms
$N_{\rm interior}$ as well as the cluster size $N$.
Using these expressions, we computed the 
concentration of the fcc motif for the two series, as indicated
by the dashed and dotted lines in Fig. 5. 
The overall trend of structure evolution is evident.
Beyond $N=147$ (three completed shells in an icosahedral structure),
the ground state essentially alternates between the two structures.
The main transition between the icosahedra-dominated regime 
at smaller sizes to the decahedra-dominated regime at larger sizes
takes place at around $N=1030$, indicated by an upward jump in 
the $c_{\rm fcc}$ curve (black solid line) on the plot. 
Windows exist on either side of the transition where the minority series 
pops in. The physical mechanism behind these transitions,
which has to do with the competition between strain and surface
energies, has already been discussed extensively in the 
literature.\cite{ref:Doye2002}

\begin{table}
\caption{\label{tab:table1}
Number of icosahedral (ico), decahedral (dec), hexagonal-close-packed (hcp)
and face-centered-cubic (fcc) coordinated atoms in the complete Mackay
icosahedra (MIC) and Marks decahedra (MD) series. The total number of 
completed shells (including surface atoms) is given by $n$.
}
\begin{ruledtabular}
\begin{tabular}{lcc}
&MIC&MD\\
\hline
$N_{\rm ico}$& 1 & 0\\
$N_{\rm dec}$& $12(n-1)$ & $2n-1$\\
$N_{\rm hcp}$& $15(n-1)(n-2)$& ${5\over 2}(n-1)(3n-2)$\\
$N_{\rm fcc}$& ${10\over 3}(n-1)(n-2)(n-3)$& ${5\over 6}n(n-1)(4n-5)$\\
$N_{\rm interior}$& $(2n-1)[{5\over 3}n(n-1)+1]$ & 
${10\over 3}n^3-{19\over 3}n+4$\\
$N$& $(2n+1)[{5\over 3}n(n+1)+1]$ & 
${10\over 3}n^3+10n^2+{11\over 3}n+1$\\
\end{tabular}
\end{ruledtabular}
\end{table}

\begin{figure}
\narrowtext
\centerline{
\epsfxsize=\linewidth
\epsfbox{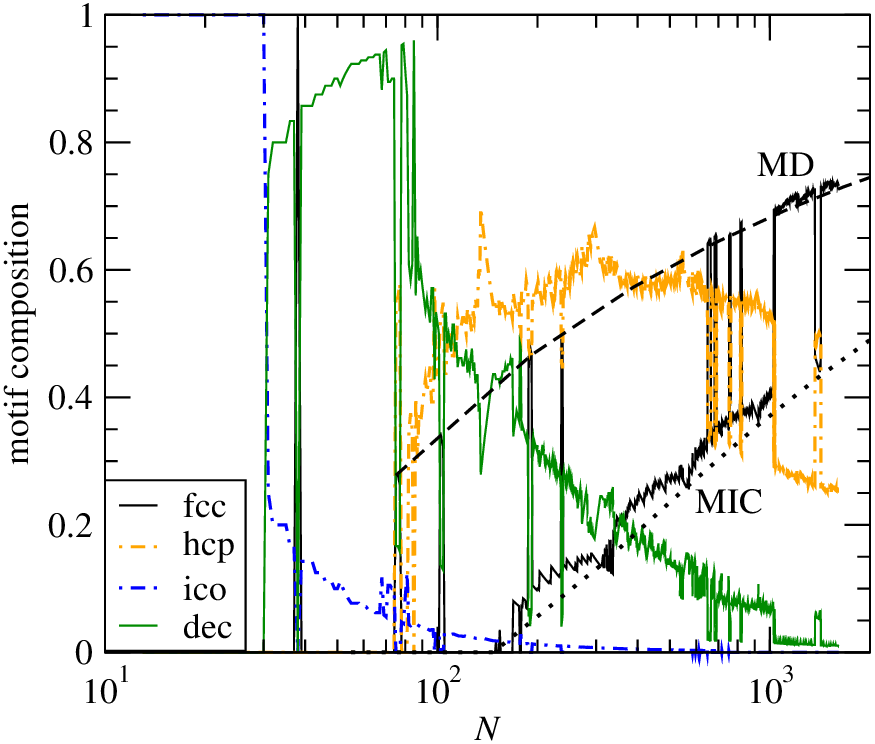}
}
\caption{(Color online) Composition profile of four coordination motifs
in the ground state LJ clusters. Dashed and dotted lines
give the theoretical values of $c_{\rm fcc}$ for complete Mackay icosahedra
and Marks decahedra, respectively.
}
\label{fig5}
\end{figure}

We have examined the spatial distribution of the coordination motifs on
large atomic clusters.
Figure 6 shows the position of ico (blue), dec (green), and hcp (orange)
atoms at $N=923$ (MIC) and $1103$ (MD).
The icosahedrally coordinated atom is only present at the center of the 
MIC cluster. The decagonally coordinated atoms are located on the 
five-fold symmetry axes in both clusters, while the hcp atoms form 
two-dimensional sheets that separate the fcc grains (not shown) at
symmetry related crystallographic orientations.

\begin{figure}
\narrowtext
\centerline{
\epsfxsize=0.9\linewidth
\epsfbox{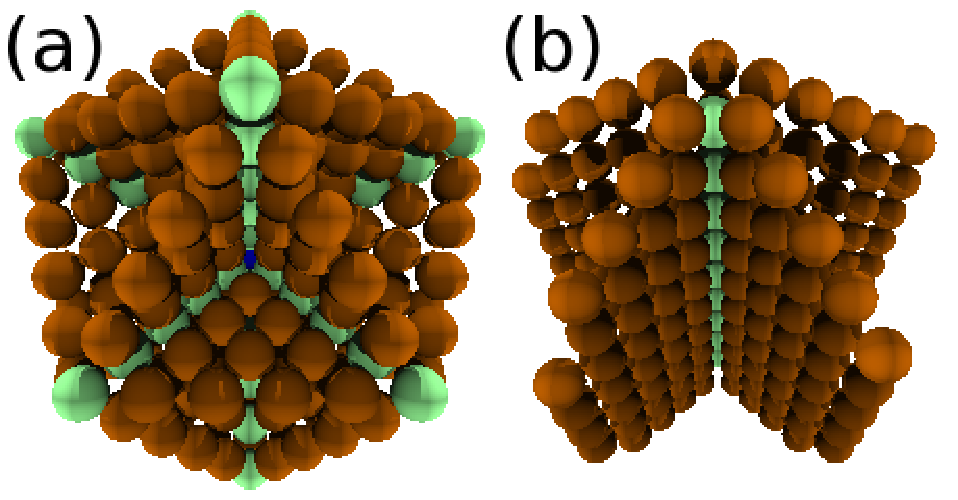}
}
\caption{(Color online) Position of ico (blue), dec (green), and 
hcp (orange) atoms in the ground state LJ cluster at size
(a) $N=923$ (Mackay icosahedron) and (b) $N=1103$ (Marks decahedron), 
showing large-scale structural organization.
}
\label{fig6}
\end{figure}

Clusters of size less than three atomic layers exhibit a greater variety
of organizational behavior due to the more subtle and delicate
balance of packing, strain, and surface effects. 
Figure 7 shows the composition profiles in this
size range for two representative classes of ground states in the CCD:
the LJ clusters discussed above and the C$_{60}$ clusters
computed using the Pacheco and Prates-Ramalho potential.\cite{ref:Doye2001}
Most of the LJ clusters in this size range has a single icosahedron
motif (open blue circle) at the core. Their structure essentially follows
that of the Mackay icosahedra series, with modifications close to the
surface layer. Notable exceptions are found at $N=38$ (fcc) and
$N=75-77$ (Marks decahedron), where the icosahedron core is missing.

The C$_{60}$ clusters behave quite differently from the LJ clusters.
The icosahedron core is present only at $N\leq 15$.
About one third of the clusters in the size range $31-100$ are cubic
(i.e., with only fcc and hcp motifs), whereas the rest contain
some form of the decahedral structure.
It has been suggested that, due to the narrow range of the molecular
potential, C$_{60}$ clusters prefer close-packed structures as compared to
the LJ clusters.\cite{ref:Doye2001,ref:Doye1995} This is in good agreement 
with the much higher percentage of fcc, hcp, and dec motifs in the 
composition profiles as shown in Fig. 7. We have also examined the different 
series of the ground state Morse clusters in the CCD that correspond 
to a broad range of softness of the pair potential.\cite{ref:Doye1996} 
Indeed, as the range of the interaction
shortens, a change from the icosahedral to decagonal and cubic structures
is observed. 

\begin{figure}
\narrowtext
\centerline{
\epsfxsize=0.9\linewidth
\epsfbox{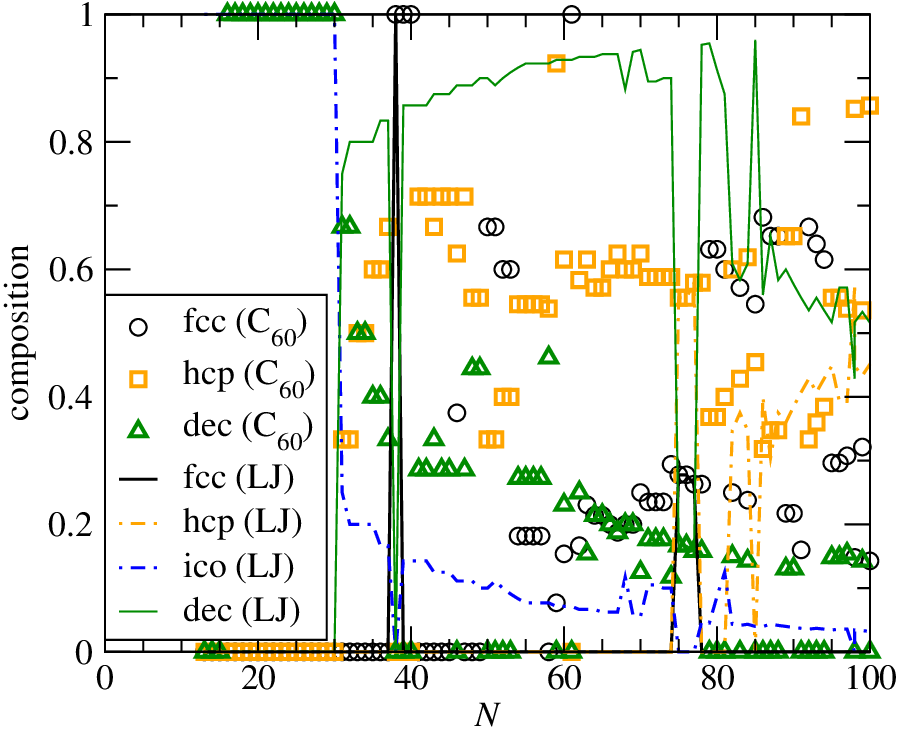}
}
\caption{(Color online) Motif composition profile for Lennard-Jones (LJ)
and C$_{60}$ clusters below $N=100$.
}
\label{fig7}
\end{figure}

The decahedral motif plays a key role in the large-scale structural
organization due to its unique pentagonal symmetry axis.
In both the Mackay icosahedra and Marks decahedra series, 
decahedra motifs stack together to form linear chains 
to propagate the (local) pentagonal symmetry all the way to the cluster 
surface. Two or more pentagonal axes can emanate from an icosahedral
motif or an incomplete icosahedron at the surface. 
Chains of decahedra have also been observed in less ordered LJ clusters
by Polak.\cite{ref:Polak03,ref:Polak08} 
The network of decahedral motifs (plus their orientation), to a great
degree, defines the overall structure of these low energy atomic clusters.

\section{Summary and conclusions}

In summary, we have presented a general method to identify re-occuring
local structural units in an atomic cluster without prior information.
The structural units, taken here to be atoms on the first coordination
shell, are represented by a set of multipole moments whose clustering 
properties form the basis for robust and efficient classification 
schemes. Applying the scheme to the ground state of the LJ and C$_{60}$ 
clusters, we reconfirmed that the fcc, hcp, ico and dec
motifs provide a complete set of the 12-atom neighborhood for all interior
atoms. For the LJ clusters, the motif compositions show a systematic 
dependence on cluster size, in agreement with the existing understanding
on the structure evolution of these clusters. 
Large-scale structural organization are readily revealed when
interior atoms are displayed according to their coordination type. 
Under this scheme, the level of complexity in structure representation 
is significantly reduced. 
This ``microstructure'' characterization can play an important role in the
development of coarse-grained models that combine
geometry and energetics to elucidate the equilibrium
and kinetic properties of medium-sized atomic clusters at low temperatures.

The study of noisy atomic clusters allows one to relate the size and shape
of the cloud surrounding each ideal shell structure in the shape space
to the actual displacement of atoms in the physical space. 
Interestingly, the noise level at which the
clouds for different coordination motifs overlap coincides well 
with the one given by the Lindemann criterion for crystal melting.
This observation offers a potentially fruitful way to analyze melting
in super-heated solids\cite{ref:jin} in terms of the local coordination
patterns and their spatial distribution.

We wish to thank Dr. X.-G. Shao for providing the LJ cluster coordinates in 
the size range 1001-1610. This work was supported by the Research Grants 
Council of the HKSAR through grant HKBU 2020/04P.

\end{document}